\documentclass[aps,prl,preprint,superscriptaddress]{revtex4}

\usepackage{rotating}
\usepackage{longtable}
\usepackage{graphicx}
\newcommand{\di}{\displaystyle}
\usepackage{here}
\begin{document}

\title{Angle-resolved photoemission spectroscopy of perovskite-type 
transition-metal oxides 
and their analyses using tight-binding band structure}

\author{H. Wadati}
\email{wadati@wyvern.phys.s.u-tokyo.ac.jp}
\affiliation{Department of Physics and Department of Complexity 
Science and Engineering, University of Tokyo, 
Kashiwa, Chiba 277-8561, Japan}

\author{T. Yoshida}
\affiliation{Department of Physics and Department of Complexity 
Science and Engineering, University of Tokyo, 
Kashiwa, Chiba 277-8561, Japan}

\author{A. Chikamatsu}
\affiliation{Department of Applied Chemistry, University of Tokyo, 
Bunkyo-ku, Tokyo 113-8656, Japan}

\author{H. Kumigashira}
\affiliation{Department of Applied Chemistry, University of Tokyo, 
Bunkyo-ku, Tokyo 113-8656, Japan}

\author{M.~Oshima}
\affiliation{Department of Applied Chemistry, University of Tokyo, 
Bunkyo-ku, Tokyo 113-8656, Japan}

\author{H. Eisaki}
\affiliation{National Institute of Advanced Industrial Science and Technology, 
Tsukuba 305-8568, Japan}

\author{Z.-X. Shen}
\affiliation{Department of Applied Physics and Stanford Synchrotron 
Radiation Laboratory, Stanford University, Stanford, California 94305,
USA}

\author{T. Mizokawa}
\affiliation{Department of Physics and Department of Complexity 
Science and Engineering, University of Tokyo, 
Kashiwa, Chiba 277-8561, Japan}

\author{A. Fujimori}
\affiliation{Department of Physics and Department of Complexity 
Science and Engineering, University of Tokyo, 
Kashiwa, Chiba 277-8561, Japan}

\date{\today}
\begin{abstract}
Nowadays it has become feasible to perform angle-resolved
photoemission spectroscopy (ARPES)
measurements of transition-metal oxides
with three-dimensional
perovskite structures
owing to the availability of high-quality
single crystals of bulk and epitaxial thin films.
In this article, we review recent experimental results
and interpretation of ARPES data using empirical 
tight-binding band-structure calculations. 
Results are presented for 
SrVO$_3$ (SVO) bulk single crystals, and
La$_{1-x}$Sr$_x$FeO$_3$ (LSFO) and
La$_{1-x}$Sr$_x$MnO$_3$ (LSMO)
thin films. 
In the case of SVO, from comparison of 
the experimental results
with calculated surface electronic structure, 
we concluded that the
obtained band dispersions reflect
the bulk electronic structure. 
The experimental band structures of LSFO and LSMO
were analyzed 
assuming the G-type antiferromagnetic
state and the ferromagnetic state, respectively. 
We also demonstrated that the intrinsic uncertainty
of the electron momentum perpendicular 
to the crystal surface is important for the interpretation 
of the ARPES results of three-dimensional materials.
\end{abstract}

\maketitle
\section{Introduction}
Perovskite-type $3d$
transition-metal (TM) oxides
have been attracting much interest in these decades
because of their intriguing physical properties,
such as metal-insulator transition (MIT), colossal magnetoresistance
(CMR), and ordering of spin, charge, and orbitals \cite{rev}.
Early photoemission studies combined with cluster-model
analyses have revealed that
their global electronic structure is characterized by 
the TM $3d$-$3d$ on-site Coulomb interaction energy denoted
by $U$, the charge-transfer energy
from the occupied O $2p$ orbitals to the empty TM $3d$ orbitals denoted
by $\Delta$, and the $p$-$d$ hybridization strength denoted 
by $(pd\sigma)$ \cite{ZSA}. If $\Delta >U$, the band gap is of
the $d$-$d$ type, and the compound is called a Mott-Hubbard-type insulator.
For $\Delta<U$, the band gap is of the $p$-$d$ type: 
the O $2p$ band is located between 
the upper and lower Hubbard bands, and 
the compound is called
a charge-transfer-type insulator.
In going from lighter to heavier $3d$ TM
elements, the $d$ level is lowered
and therefore $\Delta$ is decreased,
whereas $U$ is increased.
Thus Ti and V oxides become 
Mott-Hubbard-type compounds and Mn, Fe, Co, Ni, and Cu
oxides charge-transfer-type compounds.

Having established the global electronic
structure picture of the transition-metal oxides with
their chemical trend, the next step is to understand
the momentum dependence of the electronic states,
that is, the ``band structure'' of the hybridized $p$-$d$
states influenced by the Coulomb interaction $U$.
Angle-resolved photoemission spectroscopy (ARPES) is
a unique and most powerful experimental technique
by which one can directly determine the band structure of
a material. However, 
ARPES studies have been largely limited to low-dimensional
materials and there have been
few studies on three-dimensional TM oxides
with cubic perovskite structures
because many of
them do not have a cleavage plane.
Recently, Yoshida {\it et al.}
succeeded in cleaving single crystals of
SrVO$_3$ (SVO) along the cubic $(100)$ surface
and performed ARPES measurements \cite{SVOARPES}.
Another way of performing ARPES studies of
such materials is the use of single-crystal thin films.
Recently, high-quality perovskite-type
oxide single-crystal thin films grown by the
pulsed laser deposition (PLD) method
have become available \cite{Izumi,Choi}, and setups have been
developed for their {\it in-situ} photoemission
measurement \cite{Horiba,Shi}. 
By using well-defined surfaces of epitaxial thin films, we
recently studied the band structures of
La$_{1-x}$Sr$_x$FeO$_3$ (LSFO) and
La$_{1-x}$Sr$_x$MnO$_3$ (LSMO),
and it has been demonstrated that {\it in-situ} ARPES
measurements on such TM oxide films are one of the
best methods to investigate the band structure of
TM oxides with three-dimensional crystal structures
\cite{Chika,LSFOARPES}.

In this paper, we report on analyses of ARPES results of
bulk single crystals of SVO \cite{SVOARPES}
and on {\it in-situ} ARPES
results of single-crystal LSFO ($x=0.4$) \cite{LSFOARPES}
and LSMO ($x=0.4$) \cite{Chika}
thin films grown on SrTiO$_3$ (001) substrates.
In the case of SVO, we have investigated the
effect of surface on the ARPES spectra
by comparing the experimental results
with calculated surface electronic structures.
The experimental band structures of LSFO and LSMO
were interpreted using an empirical tight-binding band-structure
calculation by assuming the G-type antiferromagnetic 
state for LSFO and the ferromagnetic state for LSMO. 
The mass enhancement compared with band-structure 
calculation is discussed in the case of SVO and LSMO. 
We also discuss the problem of intrinsic uncertainty in
determining the momentum of electrons in the solid perpendicular
to the crystal surface \cite{3DARPES}.
\section{Theoretical analyses}
\subsection{Tight-binding band-structure calculation}
In the case of SVO, we adopted 
the simplest tight-binding (TB) model. 
In this simplest model, the 
$t_{2g}$ bands are represented by a TB 
Hamiltonian with diagonal elements of the 
form 
\begin{eqnarray}
 H_{xy,xy} (\mbox{\boldmath$k$}) & = & e_d+t_0(c_x+c_y)+t_1c_xc_y
+[t_2+t_3(c_x+c_y)+t_4c_xc_y]c_z,
\end{eqnarray}
where $c_i=2\cos (k_ia)$ ($i=x$, $y$, $z$) and $a$ 
is the lattice constant \cite{Liebsch}. 
Cyclic permutations yield 
$H_{yz,yz}(\mbox{\boldmath$k$})$ and 
$H_{zx,zx}(\mbox{\boldmath$k$})$. 
The $t_i$ denote effective 
hopping integrals arising from the V-O-V hybridization. 
$t_{0,2}$, $t_{1,3}$, and $t_4$ specify the 
interaction between first, second, and third neighbors, 
respectively, and $|t_0|\gg|t_1|$, $|t_2|$, $|t_3|$, 
$|t_4|$. 
Off-diagonal elements 
which vanish at high-symmetry points are 
small and are neglected here. 

In the case of LSFO and LSMO, 
we performed TB band-structure calculations explicitly including 
oxygen $2p$ orbitals 
for the three-dimensional perovskite structure in the 
antiferromagnetic (AF) and ferromagnetic (FM) states. 
TB calculations for the paramagnetic (PM) state 
were already reported in Refs.~\cite{Kahn,ReO3,STO}. 
The size of the matrix for the latter
calculation was $14\times 14$
(TM $3d$ orbitals: 5, oxygen $2p$ orbitals:
$3\times 3 =9$).
As for LSFO, we performed the calculation assuming the
G-type AF state, where all nearest-neighbor Fe atoms have
antiparallel spins. The effect of the G-type
antiferromagnetism was treated phenomenologically
by assuming an energy difference $\Delta E$
between the spin-up and spin-down Fe sites. As for LSMO,
we performed the calculation assuming the FM state.
The effect of ferromagnetism was again treated phenomenologically
by assuming an energy difference $\Delta E$
between the spin-up and spin-down Mn sites.
The size of the matrix to be diagonalized was $28\times 28$
for LSFO (G-type AF), and $14\times 14$ for LSMO (FM).
Parameters to be adjusted are $\Delta E$,
the energy difference between the TM $3d$ level
and the O $2p$ level $\epsilon_d-\epsilon_p$,
and Slater-Koster parameters $(pd\sigma)$, $(pd\pi)$,
$(pp\sigma)$, and $(pp\pi)$ \cite{JC}. 
Here, $\epsilon_d$ is the average of the majority-spin 
$d$-level $\epsilon_d-\Delta E/2$ and the 
minority-spin $d$-level $\epsilon_d+\Delta E/2$. 
The ratio $(pd\sigma)/(pd\pi)$ was fixed at $-2.2$ \cite{WA,STO1},
and $(pp\sigma)$ and $(pp\pi)$ were fixed at $0.60$ eV and $-0.15$ eV,
respectively. $(pd\sigma)$ is expected to be 
in the range of $-(1.4-1.9)$ eV for LSFO and near $-1.8$ eV for LSMO
from a configuration-interaction (CI) cluster-model calculation 
reported in Refs.~\cite{Bocquet2,Wadati}.
Crystal-field splitting $10Dq$ of 0.41 eV was taken from
Ref.~\cite{Wadati}. Crystal-field splitting of 
O $2p$ orbitals is not taken into account. 
The TB basis functions used in the calculations of
LSFO and LSMO are given in Tables~\ref{symbol1} and
\ref{symbol3}, and
the matrix elements of the Hamiltonians for LSFO and LSMO are
given in Tables~\ref{symbol2} and \ref{symbol4}, respectively.

\begin{figure}
\begin{center}
\includegraphics[width=14cm]{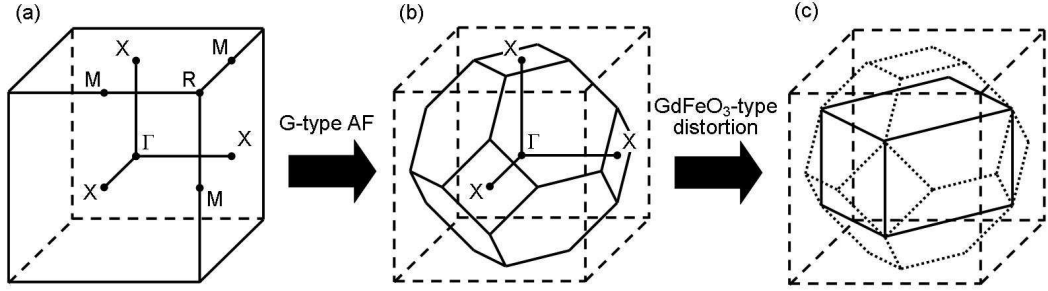}
\caption{\label{BZ} 
Brillouin zones of the perovskite structure.
(a)PM and FM states. (b)G-type AF state. Broken lines represent
the BZ of the PM state. (c)G-type AF state with
GdFeO$_3$-type distortion. Broken lines and dotted lines represent
the BZs of the PM state and the G-type AF state, respectively.}
\end{center}
\end{figure}

Figure \ref{BZ} shows the Brillouin zones (BZs) for the 
perovskite structure in 
the PM and FM states [(a)], 
the G-type AF state [(b)], and 
the G-type AF state with GdFeO$_3$-type distortion [(c)]. 
The BZ of the FM state is the same as that of the PM state. 
The BZ of the G-type AF state is that of the 
face-centered cubic (fcc) lattice. 
In this paper, we will neglect the effect of 
the GdFeO$_3$-type distortion and will only consider the effect of
magnetic ordering.

\begin{figure}
\begin{center}
\includegraphics[width=14cm]{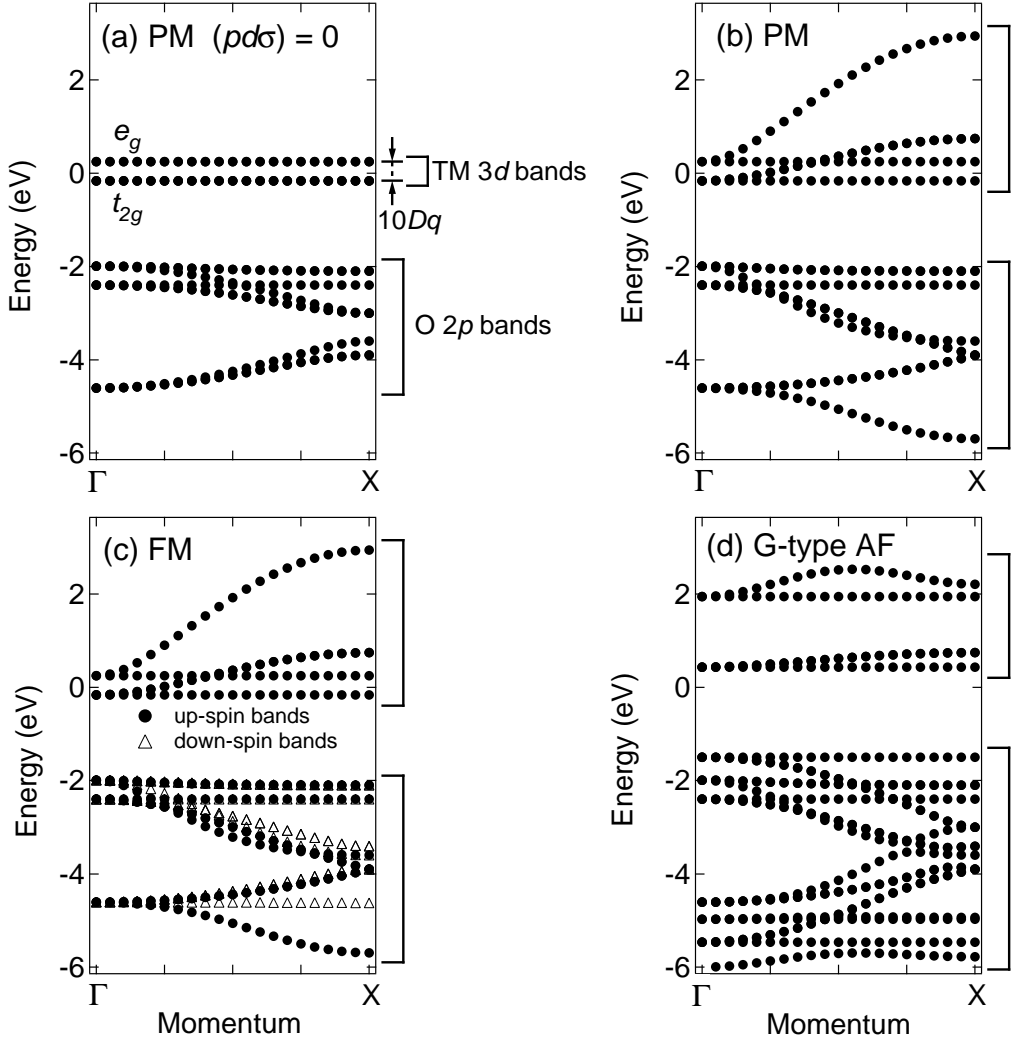}
\caption{\label{TBFAF} Band structures obtained 
by TB calculation. 
$\epsilon_d-\epsilon_p-\Delta E/2=3$ eV has been fixed. 
(a)PM state with only $p$-$p$ interaction included ($(pd\sigma)=0$).
(b)PM state with $(pd\sigma)=-2.0$ eV. 
(c)FM state with $\Delta E=5.0$ eV. 
(d)G-type AF state with $\Delta E=5.0$ eV.}
\end{center}
\end{figure}

Figure \ref{TBFAF} shows schematic band structures 
along the $\Gamma$-X line obtained 
by TB calculation. Here, we have fixed 
$\epsilon_d-\epsilon_p-\Delta E/2$ at 3 eV. 
Panel (a) shows the band structure of the PM state 
calculated including only $p$-$p$ interactions, namely, 
without $p$-$d$ interaction ($(pd\sigma)=0$). 
There are significant dispersions in the O $2p$ bands caused by 
the $p$-$p$ interaction, whereas there are no dispersions in
the $e_g$ and $t_{2g}$ bands, where there exists only 
the crystal-field splitting of $10Dq$. 
Panel (b) shows the band structure of the PM state 
calculated 
with $(pd\sigma)=-2.0$ eV. The $e_g$ and $t_{2g}$ bands
now become dispersive due to the $p$-$d$ interaction. 
Panels (c) and (d) show the band structure of 
the FM and G-type AF states, respectively. 
The values of $\Delta E$ were taken as 5.0 eV in both cases. 
In the FM state, since $\epsilon_d-\epsilon_p-\Delta E/2$ 
is fixed, 
the up-spin (majority-spin) $e_g$ and $t_{2g}$ bands 
do not change 
appreciably compared with those in the PM state, 
whereas the 
down-spin (minority-spin) $e_g$ and $t_{2g}$bands 
are pushed up by $\Delta E$ 
(out of the range of Fig.~\ref{TBFAF} (c)). 
In the G-type AF state, band dispersions along the 
R-M line are folded onto the $\Gamma$-X line, and 
hybridization between these bands 
causes a maximum in the $e_g$ bands 
in the middle of $\Gamma$ and X. 
\subsection{$k_z$ broadening}
In ARPES measurements, there is an intrinsic
uncertainty in the momentum of electrons 
in the solid perpendicular to the crystal surface due to 
a finite escape depth $\lambda$ of photoelectrons \cite{3DARPES}.
Here, the $z$-axis is taken as the direction perpendicular to the
surface toward the vacuum side. 
Inside the solid ($z<0$) 
the wave function of an emitted photoelectron 
with a wave number $k_{z0}$ is approximated by 
\begin{equation}
\psi(z)=\frac{1}{\sqrt{\lambda}}\exp(ik_{z0}z)\exp(z/2\lambda ),
\label{1}
\end{equation}
where $\lambda$
denotes the escape depth of a photoelectron.
$|\psi(z)|^2=\exp(z/\lambda)/\lambda$ shows an exponential
decay of photoelectrons within the solid. 
The Fourier transform of Eq. (\ref{1}), 
$\phi(k_z)$, is
\begin{equation}
\phi(k_z) \propto \frac{1}{i(k_z-k_{z0})-1/2\lambda}.
\end{equation}
and $|\phi(k_z)|^2$ becomes proportional to
\begin{equation}
L(k_z)=\frac{1}{2\pi\lambda}\frac{1}{(k_z-k_{z0})^2+(1/2\lambda)^2},
\label{kz}
\end{equation}
which means that the intrinsic
uncertainty of $k_z$ is given by the Lorentzian function with the full
width at half maximum of $1/\lambda$.

\begin{figure}
\begin{center}
\includegraphics[width=8cm]{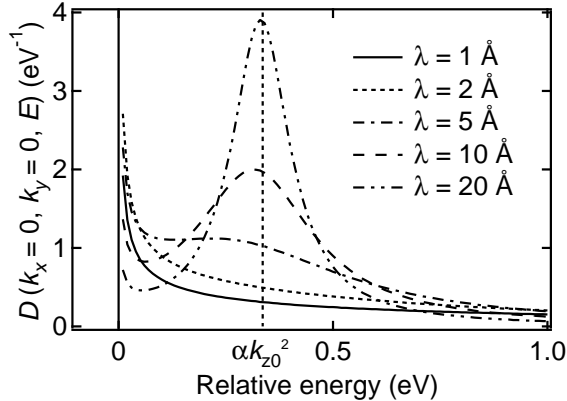}
\caption{\label{calcDOS} 
ARPES spectrum $D(k_x,k_y,E)$ with $k_x=k_y=0$ for various values of escape depths
of photoelectrons $\lambda$.}
\end{center}
\end{figure}

The effect of the $k_z$ broadening is very important for
the interpretation of ARPES results. For example, we consider an
isotropic band with the bottom at the $\Gamma$ point, expressed as
\begin{equation}
E=\alpha(k_x^2+k_y^2+k_z^2).
\end{equation}
If we measure photoelectrons with momentum
$\overrightarrow{K}=(K_x.K_y,K_z)$, where $K_z$ is related to $k_{z0}$
through the momentum conservation $k_x=K_x$ 
and $k_y=K_y$ and the energy conservation, 
the obtained spectrum becomes,
\begin{equation}
 D(k_x,k_y,E)=\di\int^{\infty}_{-\infty}dk_zL(k_z)\delta (E-\alpha
({k_x}^2+{k_y}^2+{k_z}^2)).
\end{equation}
From Eq.~(\ref{kz}) and using the identity
\begin{equation}
\delta (x-\alpha {k_z}^2)=\frac{1}{2\sqrt{\alpha x}}
\left[\delta\left(k_z-\sqrt{\frac{x}{\alpha}}\right)+
\delta\left(k_z+\sqrt{\frac{x}{\alpha}}\right)\right],
\end{equation}
we obtain
\begin{equation}
\begin{array}{c}
D(k_x,k_y,E)=\displaystyle\frac{1}{4\pi\lambda\sqrt{\alpha
\{E-\alpha({k_x}^2+{k_y}^2)\}}}\times\\
\left[\displaystyle\frac{1}{\left(\sqrt{\{E-\alpha({k_x}^2+{k_y}^2)\}/
\alpha}-k_{z0}\right)^2+(1/2\lambda)^2}\right. \\
+\left.\displaystyle\frac{1}{\left(\sqrt{\{E-\alpha({k_x}^2+{k_y}^2)\}/
\alpha}+k_{z0}\right)^2+(1/2\lambda)^2}\right].
\end{array}
\end{equation}
We plot the function $D(k_x,k_y,E)$ with $k_x=k_y=0$ for various values
of $\lambda$ in Fig.~\ref{calcDOS}. Here, we have assumed that 
$\alpha =8$ eV$\cdot \mbox\AA^2$, which
corresponds to $m^{\ast}\sim 0.5m_e$ ($m^{\ast}$ is an effective
band mass and $m_e$ is the electron mass in vacuum), and
$k_{z0}=\pi/4c$, where $c=3.830\ \mbox{\AA}$, the 
out-of-plane lattice constant of 
LSMO ($x=0.4$) thin films. One can see 
that in bulk-sensitive measurements using high-energy
photons, like $\lambda =20\ \mbox{\AA}$,
there is a peak at $E=\alpha {k_{z0}}^2$.
In contrast, in surface-sensitive measurements using
low-energy photons,
there is no peak at $\alpha {k_{z0}}^2$ but $D(E)$ is
peaked around
$E\sim 0$. The density of states at the
$\Gamma$ point is very large because
the band has a van Hove singularity at this point.
Since bands usually have van Hove singularities along a
high-symmetry line, the effect of $k_z$ broadening is
very important when we trace near the high-symmetry line
using low energy photons. Our photon energies used in the ARPES
measurements were in the range of $20-100$ eV and therefore 
$\lambda \sim 5 \mbox{\AA}$, 
meaning that considering the $k_z$-broadening 
effect is crucial in 
the interpretation of ARPES results.

\section{Results and discussion}
\subsection{SrVO$_3$}
SVO is an ideal system 
to study the fundamental physics of electron correlation
because it is a prototypical
Mott-Hubbard-type with $d^1$ electron configuration. 
For the bandwidth-control system 
Ca$_{1-x}$Sr$_x$VO$_3$ (CSVO), 
the existence and the extent 
of effective mass 
enhancement and spectral weight transfer
have been quite controversial. 
Early photoemission results 
have shown that, with decreasing $x$, i.e., 
with decreasing bandwidth, 
spectral weight is transferred from the coherent part
to the incoherent part \cite{Inoue} in a more dramatic way than the
enhancement of the electronic specific heats $\gamma$ \cite{Inoue2}.
In contrast, in a recent bulk-sensitive
photoemission study using soft x rays it was claimed
that there was no appreciable spectral weight transfer between
SrVO$_3$ and CaVO$_3$ \cite{Sekiyama}.

ARPES measurements on SVO were performed at beamline 5-4 of
Stanford Synchrotron Radiation Laboratory (SSRL). Bulk single
crystals of SVO were grown using the traveling-solvant floating
zone method. Samples were first aligned {\it ex situ} using Laue
diffraction, cleaved {\it in situ} along the cubic (100) surface
at a temperature of 15 K and measured at the same temperature.
Details of the experimental conditions of ARPES are described in
Ref. \cite{SVOARPES}.

\begin{figure}
\includegraphics[width=12cm]{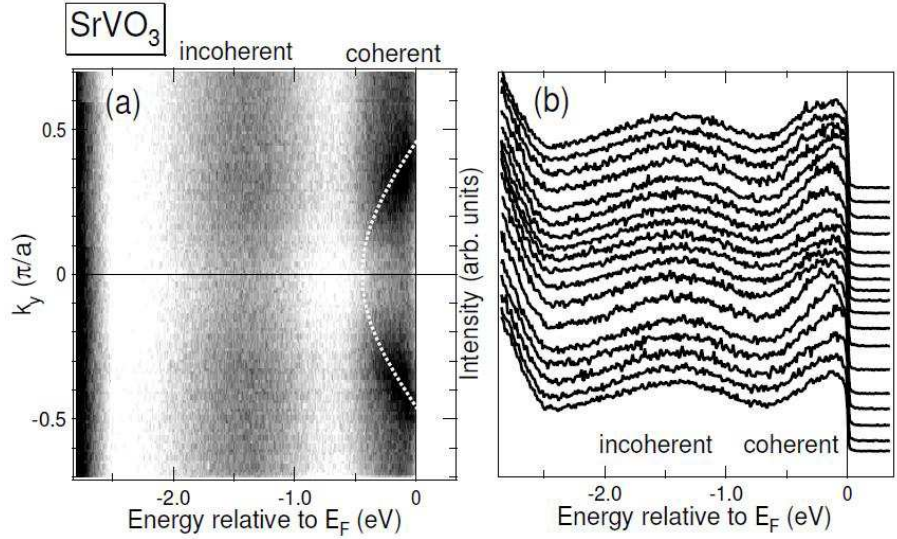}
\caption{\label{incoherent} ARPES spectra of SrVO$_3$ along
momentum cut 2 in the Brillouin zone shown in Fig. \ref{TBFS}. (a)
Intensity plot in the $E$-$k_y$ plane. Dispersive feature within
$\sim$0.7 eV of $E_F$ is the coherent part, while a broad feature
around -1.5 eV is the incoherent part. (b) Energy distribution
curves (EDC's). }
\end{figure}

An example of ARPES spectra for SVO near $E_F$ are shown in Fig.
\ref{incoherent}. The coherent part within $\sim$ 0.7 eV of $E_F$
shows a parabolic energy dispersion, consistent with the view that
the coherent part corresponds to the calculated band structure. On
the other hand, the incoherent part centered at $\sim$ -1.5 eV
does not show an appreciable dispersion but shows a slight
modulation of the intensity, which coincides with that of the
coherent part, as seen in Fig \ref{incoherent}(a). To compare with
the band-structure calculation, we shall focus on the dispersive
feature of the coherent part.

\begin{figure}
\includegraphics[width=12cm]{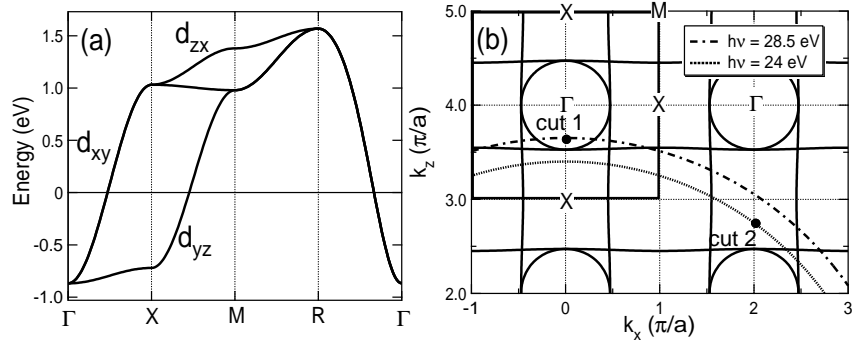}
\caption{\label{TBFS} Band structure and the Fermi surface of
SrVO$_3$. (a) Band dispersions obtained by local-density
approximation (LDA) band-structure calculation \cite{Takegahara}.
(b)$k_y$=0 cross-sectional view of the Fermi surfaces and the
momentum loci corresponding to the mapping in Fig.
\ref{nkMapping}.}
\end{figure}

\begin{figure}
\includegraphics[width=10cm]{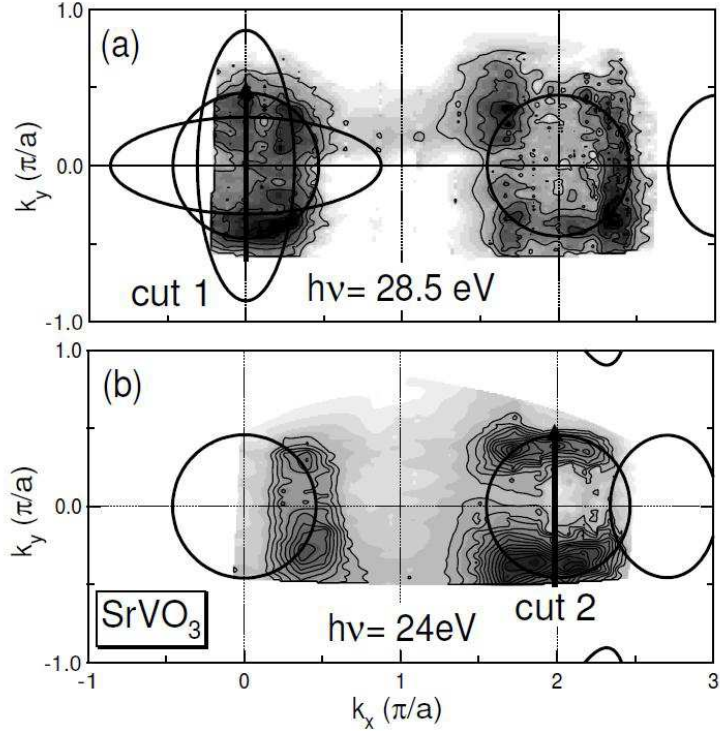}
\caption{\label{nkMapping} Spectral weight mapping at $E_F$ for
SrVO$_3$. (a) Mapping with $h\nu$=28.5eV. (b) Mapping with $h\nu$=
24 eV. Note that the mapping is projection on the $k_x$-$k_y$
plane.}
\end{figure}

The band dispersions obtained by local-density approximation (LDA)
band-calculation \cite{Takegahara} are shown in Fig. \ref{TBFS}
(a). The hopping parameters of the TB model (1) have been obtained
by fitting the TB bands to the results of LDA band-structure
calculation \cite{Takegahara}. Figure \ref{TBFS} (b) illustrates
the $k_y$=0 cross-sectional view of the Fermi surfaces expressed
by the TB model. The three bands correspond to the  $d_{xy}$,
$d_{yz}$ and $d_{zx}$ orbitals of V and scarcely hybridize with
each other. Therefore, these dispersions are almost
two-dimensional and form three cylindrical Fermi surfaces
intersecting each other around the $\Gamma$ point, although the
crystal structure is three-dimensional. The loci of electron
momenta in the $k_x$-$k_z$ plane for constant photon energies
$h\nu$= 24 eV and 28.5 eV are plotted in Fig. \ref{TBFS}(b). Here,
the inner potential of 10 eV has been assumed.

\begin{figure}
\includegraphics[width=12 cm]{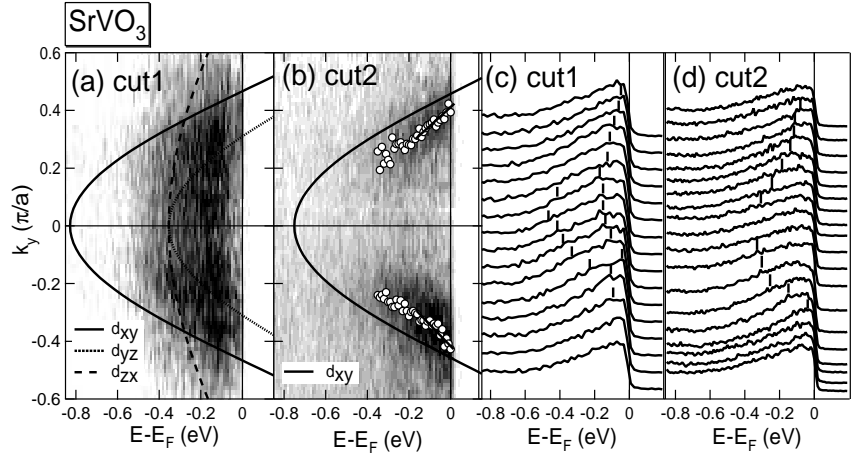}% Here is how to import EPS art
\caption{\label{EDC_Ek} ARPES spectra of SrVO$_3$ for cuts 1 and 2
in Fig. \ref{TBFS}. Panels (a) and (b) show intensity plots in
$E$-$k$ space for cuts 1 and cut 2, respectively.
Angle-independent backgrounds have been subtracted. Calculated TB
band are also superimposed. Circles in (b) are peak positions of
MDC's and represents band dispersion. Panels (c) and (d) are EDC's
corresponding to panels (a) and (b), respectively. Vertical bars
are guides to the eye indicating the positions of the dispersive
features.}
\end{figure}

Figure \ref{nkMapping} (a) and (b) shows spectral weight mapping
in the $k_x$-$k_y$ space at $E_F$ on the momentum loci for the
photon energies $h\nu$=28.5 and 24 eV, respectively. The spectral
weight for both photon energies are largely confined within the
cylindrical Fermi surfaces extended in the $k_z$ direction.
According to Fig.\ref{TBFS}(b), cut 2 corresponds to momenta near
the X point [$\mathbf{k}=(1,0,0)$]. Therefore, the observed
spectral weight comes only from one cylindrical Fermi surface
referred to as the $\gamma$ sheet \cite{inoueDHVA}, which arises
from the $d_{xy}$ orbital. On the other hand, the spectral weight
in the first BZ for $h\nu$ =28.5 eV (cut 1) is enclosed by the
three cylindrical Fermi surfaces \cite{inoueDHVA}. Indeed, EDC's
along cut 1 [Fig. \ref{EDC_Ek} (c)] show complicated features
suggestive of two energy dispersions.

%EDC's
Figure \ref{EDC_Ek}(a) and (b) show intensity plots in the
energy-momentum ($E$-$k$) space corresponding to each cut in Fig.
\ref{TBFS}. The $E$-$k_y$ plots are compared with the energy
dispersions expected from the LDA calculation \cite{Takegahara}.
Figure \ref{EDC_Ek}(c) and (d) show EDC's corresponding to panels
(a) and (b), respectively. The EDC's of cut 1 in panel (c) show
parabolic energy dispersions with the bottom at $\sim$-0.5 eV,
shallower than that expected from the LDA calculation $\sim$ -0.9
eV. Since the momenta of cut 1 are within the $\alpha$ and $\beta$
sheets of the Fermi surface, there are two dispersions. These
dispersions may correspond to the dispersions of the $d_{xy}$ and
$d_{yz}$ bands. The dispersion of $d_{zx}$ is not clearly observed
in the present plot, probably due to the effect of transition
matrix-elements.

For cut 2 [Fig. \ref{EDC_Ek} (b) and (d)], since it does not
intersect the $d_{yz}$ and $d_{zx}$ bands, only the $d_{xy}$ band
are observed. In panel (b), we have derived the band dispersion
for cut 2 from MDC peak positions indicated by circles. By
comparing the Fermi velocity of the LDA band structure and that of
the present experiment, one can see an overall band narrowing in
the measured band dispersion. Near $E_F$, we obtain the mass
enhancement factor of $m^*/m_b \sim 1.8\pm 0.2$, which is close to
the value $m^*/m_b= 1.98$ obtained from the specific heat
coefficient $\gamma$ \cite{Inoue2}. Since the LDA calculation
predicted that the $d_{xy}$ band is highly two-dimensional and
nearly isotropic within the $k_x$-$k_y$ plane, a similar mass
enhancement is expected on the entire Fermi surface.

%Surface states
It has been pointed out that the photoemission spectra of SVO
taken at low photon energies have a significant amount of
contributions from surface states, particularly in the incoherent
part \cite{Sekiyama,Maiti}. In order to distinguish bulk from
surface contributions in the present spectra, we consider the
character of surface states created by the discontinuity of the
potential at the surfaces. According to a tight-binding Green's
function formalism for a semi-infinite system \cite{Kalstein,
Liebsch}, surface-projected density of states of surface-parallel
$d_{xy}$ band and surface-perpendicular $d_{yz, zx}$ band have
different nature. The surface $d_{xy}$ band does not show an
appreciable change from the bulk band because it has the
two-dimensional character within the $x$-$y$ plane. In contrast,
spectral weight of the surface $d_{yz, zx}$ states is
redistributed between the bulk $d_{yz, zx}$ band and the Fermi
level \cite{SVOARPES}, because $k_z$ is no longer a good quantum
number. This leads to the absence of clear dispersions of the
surface $d_{yz, zx}$ bands in ARPES as demonstrated in Ref. 
\cite{SVOARPES}. Accordingly, the observed dispersion of the
$d_{xy}$ band should represent that of the bulk $d_{xy}$ band.

\subsection{La$_{1-x}$Sr$_x$FeO$_3$}
LSFO
undergoes a pronounced charge disproportionation and an associated MIT
around $x\sim 2/3$ \cite{Takano}.
One striking feature of LSFO is that the insulating
phase is unusually wide in the phase diagram
($0<x<0.5$ at room temperature and even
$0<x<0.7$ at low temperatures) \cite{Matsuno}.
In a previous photoemission study, the gap at the Fermi level ($E_F$)
was observed for all compositions for $0\le x\le 0.67$ \cite{Wadati},
consistent with the wide insulating region of this system. 

ARPES measurements on LSFO
were carried out using a photoemission spectroscopy
(PES) system combined with a laser molecular beam epitaxy (MBE)
chamber, which was installed at beamline BL-1C of Photon
Factory, KEK \cite{Horiba}.
The preparation and characterization of the LSFO films
are described in Ref.~\cite{Wadati}. By low energy 
electron diffraction (LEED), 
sharp $1\times 1$ spots were observed with no
sign of surface reconstruction,
confirming the well-ordered surfaces of the films.
Details of the experimental conditions are described in
Ref.~\cite{LSFOARPES}.

\begin{figure}
\begin{center}
\includegraphics[width=13cm]{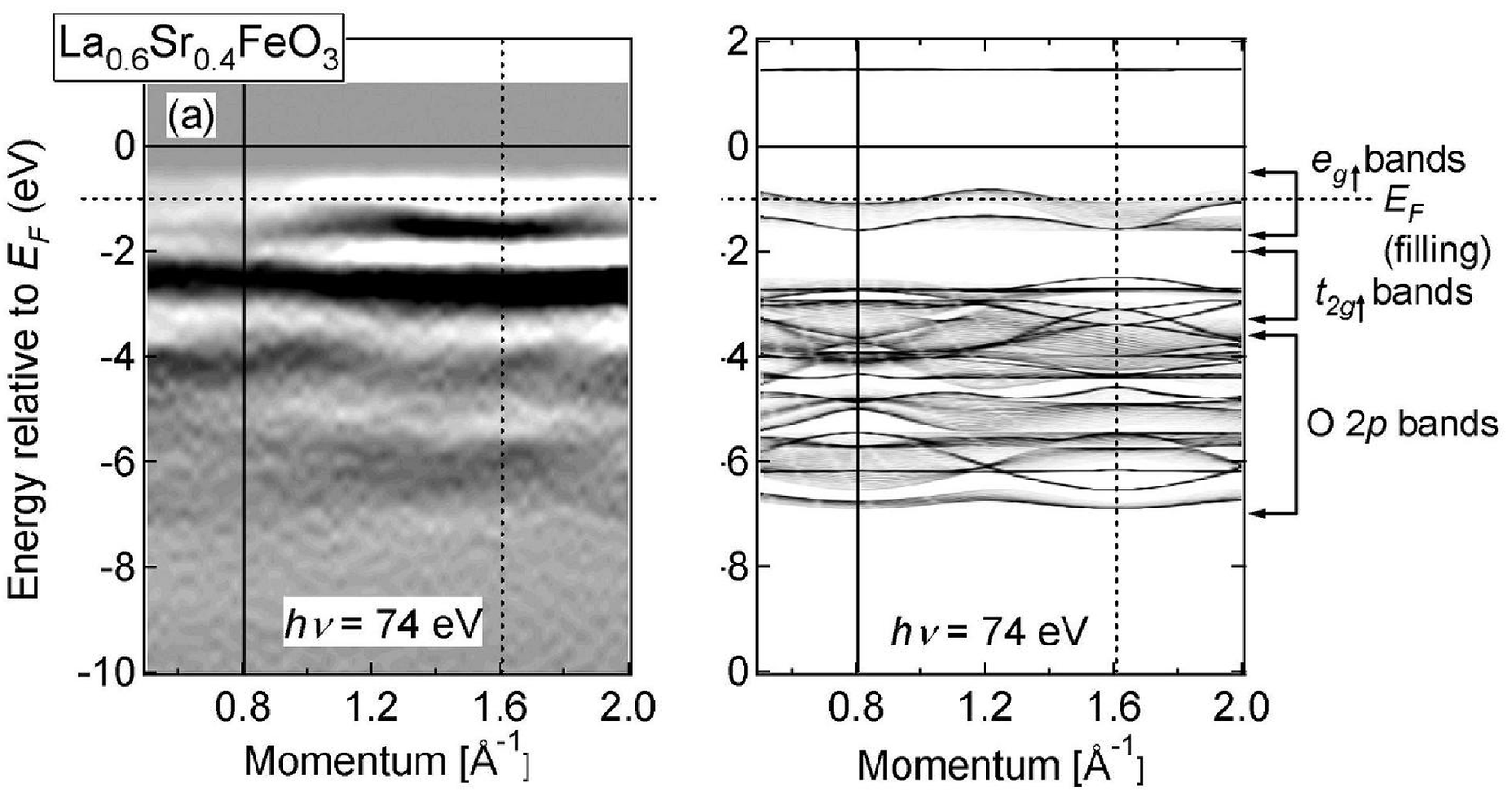}
\includegraphics[width=8cm]{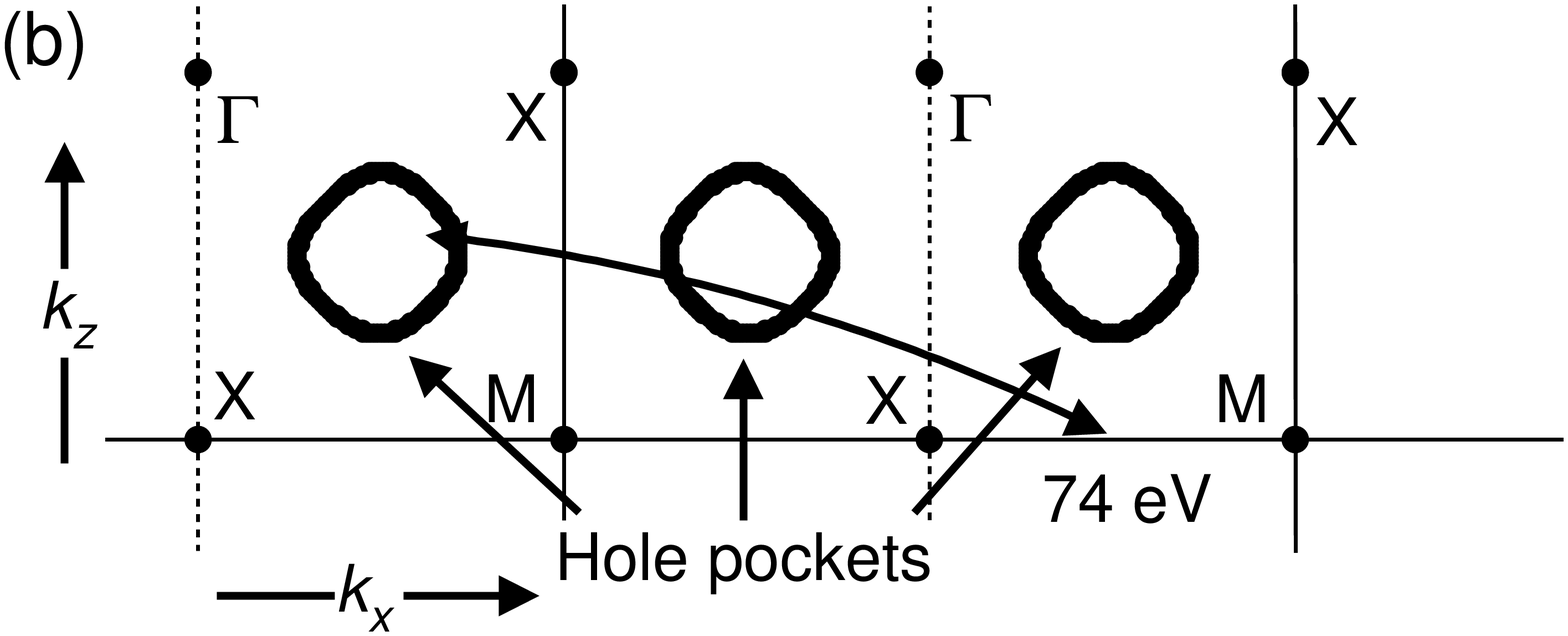}
\caption{\label{tight2} 
Comparison of ARPES spectra of an LSFO thin film
taken at $h\nu = 74$ eV with TB calculation.
The left panel of (a) shows the experimental band structure
(dark parts correspond to energy bands) and
the right panel of (a) shows the result of TB calculation.
The traces in $k$-space and calculated hole pockets are shown in (b).}
\end{center}
\end{figure}

The left panel of Fig.~\ref{tight2} (a) shows a gray-scale plot
of the experimental band structure obtained by ARPES measurements
of an LSFO ($x=0.4$) thin film with the photon energy of
74 eV \cite{LSFOARPES}. Here, the second derivatives of
the EDC's 
are plotted on the gray scale, where dark
parts correspond to energy bands.
In Fig.~\ref{tight2} (b), hole pockets obtained
by a TB calculation (described below) are
also shown. The trace 
obtained for $h\nu = 74$ eV crosses the calculated hole pocket.
The structure at $-1.3$ eV
shows a significant dispersion,
and is assigned to the Fe $3d$ majority-spin $e_g$ bands.
The structure at $-2.4$ eV shows a weaker dispersion
than the $e_g$ bands, and is assigned to the Fe $3d$
$t_{2g}$ majority-spin bands.
The structures at $-(4-7)$ eV are assigned to the O $2p$ bands.
The $e_g$ bands do not cross $E_F$,
consistent with the insulating behavior and the persistence
of the gap observed in the angle-integrated PES (AIPES) spectra
\cite{Wadati}.

The best fit of TB
calculation [right panel of 
Fig.~\ref{tight2} (a)]
to the observed band dispersions and the
optical gap of 2.1 eV of LaFeO$_3$ (LFO) \cite{opt1} 
has been obtained with $\epsilon_d-\epsilon_p$ = 2.65 eV,
$(pd\sigma) = -1.5$ eV, and $\Delta E = 5.3$ eV.
We also considered the effect of $k_z$ broadening
given by a Lorentzian function as described in Sec.~2.2.
Here we have used the value $\lambda =5\ \mbox{\AA}$.
$\Delta k_z \sim 1/\lambda$ $(\sim 0.2\ \mbox{\AA}^{-1})$ is approximately
10 \% of the Brillouin zone ($2\pi/a$).
Here, it should be noted that $\epsilon_d-\epsilon_p$, $(pd\sigma)$
and $\Delta E$ primarily determine the Fe $3d$ $-$ O $2p$ band positions,
their dispersions, and the optical band gap, respectively.
The dispersions of the $e_g$ bands were successfully reproduced. The
very weak dispersions of the $t_{2g}$ bands and the width of the O $2p$ bands
were also well reproduced by this calculation. The energy position of
the calculated $E_F$ determined from the filling of electrons 
was not in agreement 
with the experimental $E_F$, however.
This discrepancy corresponds to the
fact that this material is insulating up to 70 \% hole doping
while the rigid-band model based on the band-structure
calculation gives the metallic state.

\subsection{La$_{1-x}$Sr$_x$MnO$_3$}
LSMO has been attracting great interest because of its 
intriguing properties, such as the CMR effect 
\cite{urushibara} and half metallic nature 
\cite{parknat}. One end member of LSMO, 
LaMnO$_3$, is an antiferromagnetic insulator, 
and hole-doping induced by the substitution of Sr 
for La produces a ferromagnetic metallic phase 
($0.17<x<0.5$) \cite{urushibara}. 

ARPES measurements on LSMO
were carried out using the
same system as in the case of LSFO. 
The preparation and characterization of the LSMO films
are described in Refs.~\cite{horibaLSMO}. The surface
structure and cleanliness of the films were checked
by LEED. Some
surface reconstruction-derived spots were observed
in addition to sharp $1\times 1$ spots,
confirming the well-ordered surfaces of the films, but
surface reconstruction is expected to affect the
obtained band dispersions.
Details of the experimental conditions are described in
Ref.~\cite{Chika}.

\begin{figure}
\begin{center}
\includegraphics[width=13cm]{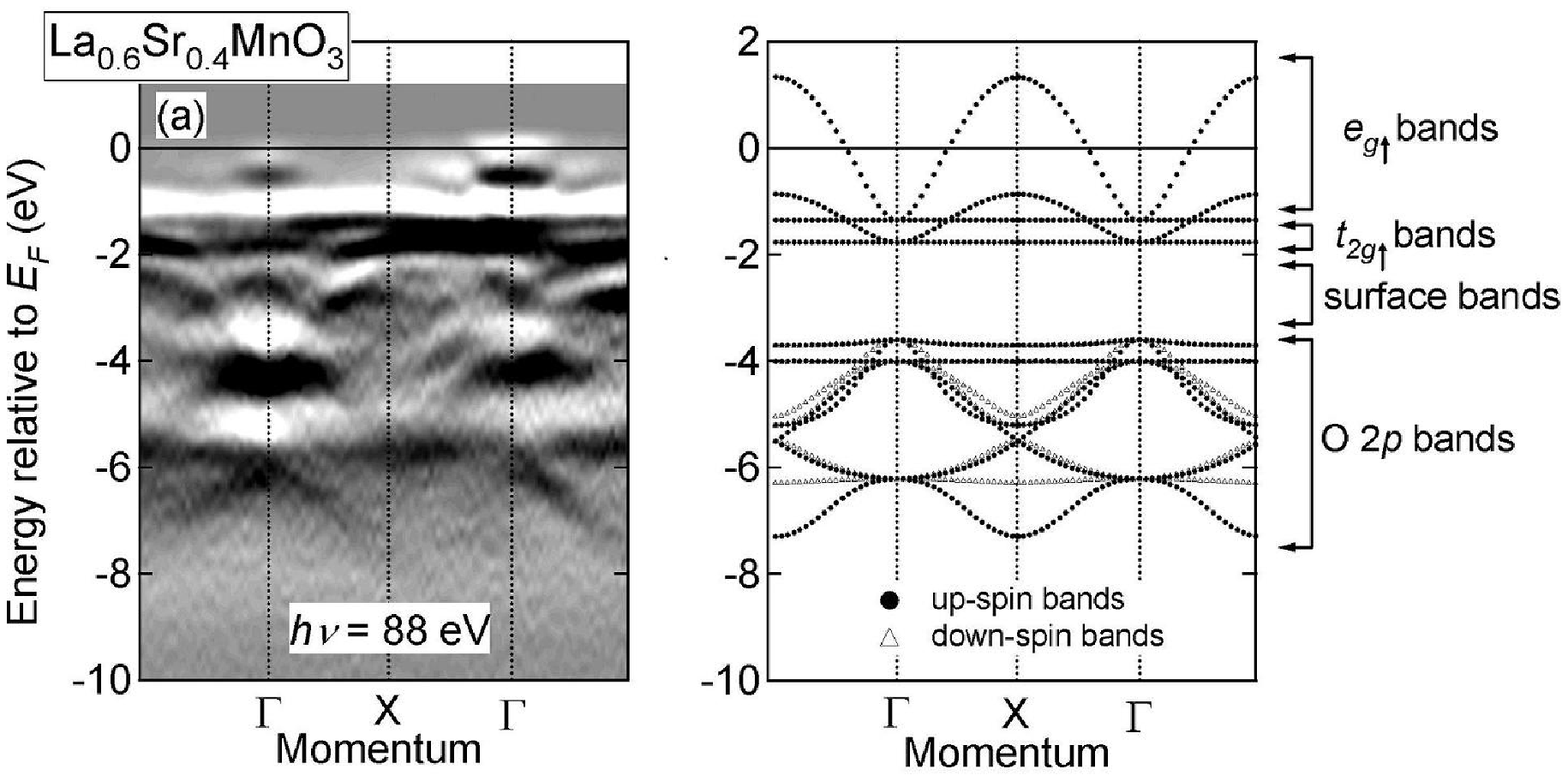}
\includegraphics[width=8cm]{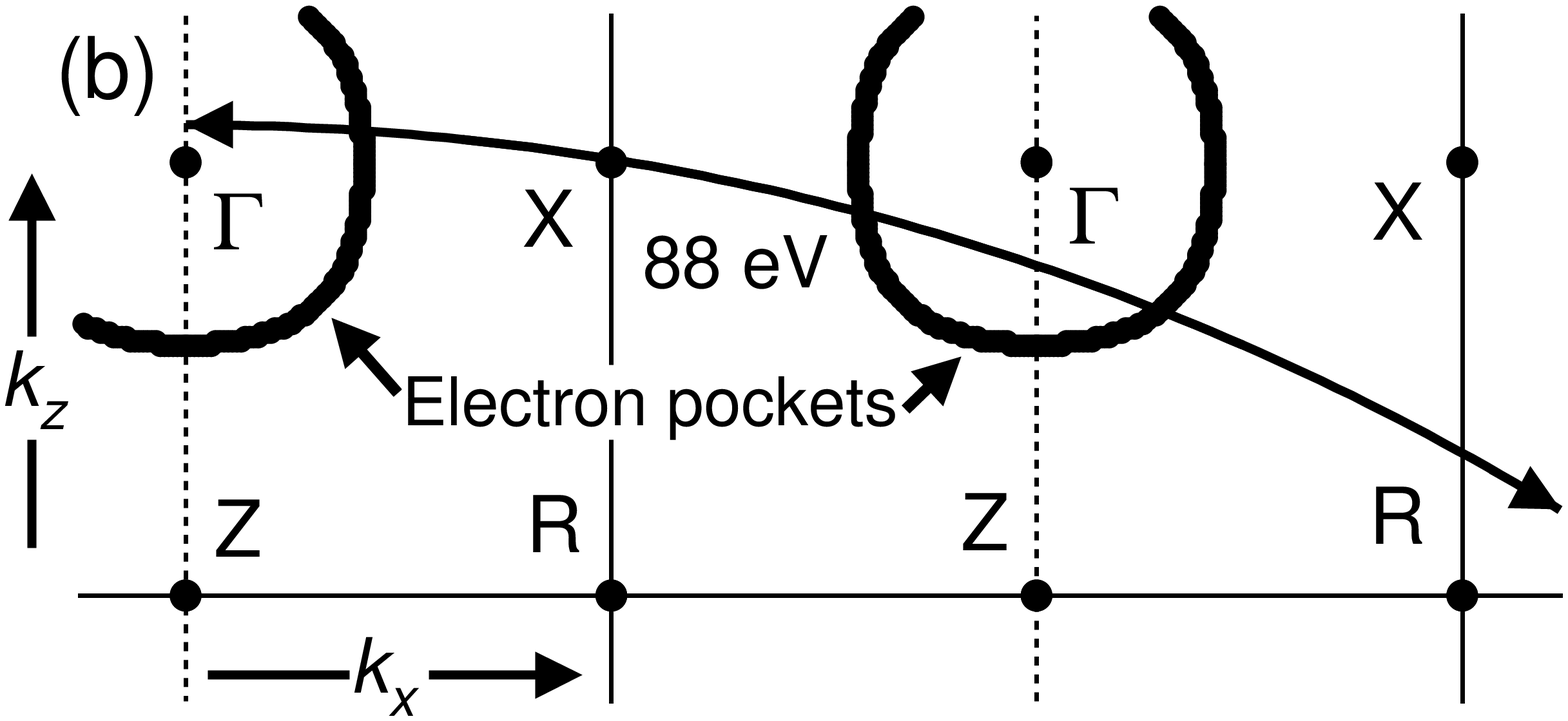}
\caption{\label{tight3} Comparison of ARPES spectra of an LSMO thin film
taken at $h\nu = 88$ eV and TB calculation.
The left panel of (a) shows the experimental band structure
(dark parts correspond to energy bands) \cite{Chika} and
the right panel of (a) shows the result of TB calculation.
The traces in $k$-space and calculated electron pockets are shown in (b).}
\end{center}
\end{figure}

The left panel of
Fig.~\ref{tight3} (a) shows a gray-scale plot of
the experimental band structure
obtained by ARPES measurements of an LSMO ($x=0.4$) thin film
with the photon energy of 88 eV \cite{Chika}.
Here, the second derivatives of the EDC's
are plotted on the gray scale, where dark
parts correspond to energy bands as in the case of LSFO 
along the trace in $k$-space shown in Fig.~\ref{tight3} (b). 
Note that this trace is approximately 
along the $\Gamma$-X direction. 
We shall assume that the trace is exactly along the 
$\Gamma$-X direction for the moment. 
The effect of the deviation from the $\Gamma$-X 
line will be discussed below. 
Here, electron pockets obtained
by a TB calculation are also shown 
in Fig.~\ref{tight3} (b). 
One of the Mn $3d$ majority-spin
$e_g$ bands at $-(0-1)$ eV
shows a significant dispersion and
crosses $E_F$, corresponding to the half metallic
behavior \cite{parknat}.
The $E_F$ crossing of the $e_g$ bands is
consistent with
the observed Fermi edge
in the AIPES spectra
\cite{horibaLSMO}.
The Mn $3d$ majority-spin
$t_{2g}$ bands at $-(1-2)$ eV
show weaker dispersions
than the $e_g$ bands.
The bands at $-(2-3)$ eV shows dispersions
with a shorter period in $k$-space than
the Brillouin zone.
This means that these bands are
affected by the
surface reconstruction detected in the LEED 
pattern. 
The O $2p$ bands are located at $-(4-7)$ eV.

The best fit of the TB 
calculation to the observed band 
dispersions [right panel of 
Fig.~\ref{tight3} (a)] 
has been obtained with $\epsilon_d-\epsilon_p = 5.3$ eV, 
$(pd\sigma) = -2.0$ eV, and $\Delta E = 4.6$ eV. 
The strong dispersions of the $e_g$ bands, 
the weak dispersions of the $t_{2g}$ bands,
and the width of the O $2p$ bands
were successfully reproduced,
although there are no bands in the calculation 
corresponding to the experimental surface bands.
The bottom of the $e_g$ bands is at $\sim -0.5$ eV in the experiment
and at $\sim -1.5$ eV in the calculation.
This difference may come from the mass 
renormalization caused by strong electron correlation in this system. 
One can estimate the mass enhencement to be 
$m^{\ast}/m_b\sim 2.5$. 

\begin{figure}
\begin{center}
\includegraphics[width=12cm]{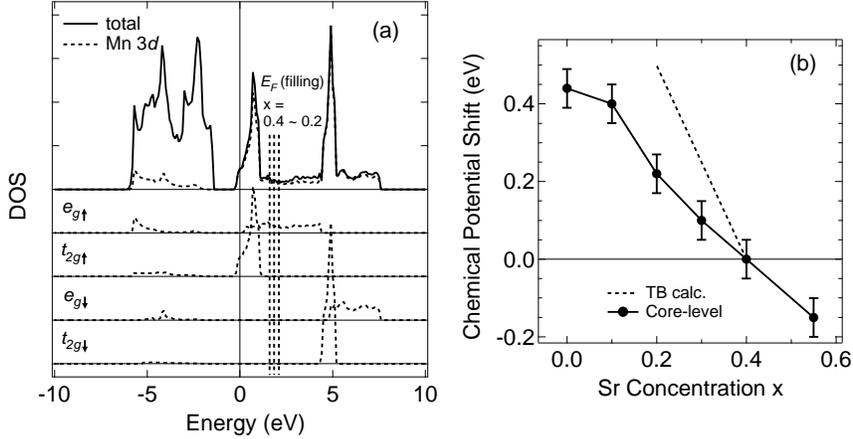}
\caption{\label{gamma} Chemical potential shift in LSMO. (a)DOS 
obtained from TB calculation.
(b)Comparison of the chemical potential shift
calculated for the rigid-band-model using TB calculation 
and that determined 
experimentally from core-level photoemission spectra
\cite{horibaLSMO}.}
\end{center}
\end{figure}

Figure \ref{gamma} (a) shows the calculated
density of states (DOS) of the FM state.
The partial DOS's for
the majority-spin Mn $e_g$,
the majority-spin Mn $t_{2g}$,
the minority-spin Mn $e_g$, and
the minority-spin Mn $t_{2g}$ orbitals
are shown in the lower panels. No band gap opens between
the majority-spin $e_g$ bands and the
minority-spin $t_{2g}$ bands for the present $\Delta E$ value.
In Fig.~\ref{gamma} (a), the $E_F$ position has been determined 
from the filling of electrons, and also calculated
DOS at $E_F$ [$D^b(E_F)$] has been obtained. 
Using the relationship
\begin{equation}
\gamma^b=\frac{\pi^2}{3}{k_B}^2D^b(E_F),
\end{equation}
we obtain the 
calculated electronic specific heat coefficient by 
$\gamma^b\sim 1$ [mJ mol$^2$ K$^{-2}$]. 
The experimental electronic specific heat coefficient 
has been reported as $\gamma \sim 3.5$
[mJ mol$^2$ K$^{-2}$] \cite{okuda}. Thus 
$m^{\ast}/m_b=\gamma/\gamma_b\sim 3.5$, in fairly good agreement
with $m^{\ast}/m_b\sim 2.5$ derived from ARPES, which means that
effective mass $m^{\ast}$ is renormalized as
\begin{equation}
m^{\ast}/m_b \sim 2.5 - 3.5.
\label{25}
\end{equation}

Figure \ref{gamma} (b) shows the comparison of the chemical
potential shift obtained from the 
TB calculation in the FM region of 
LSMO ($0.2\le x \le 0.4$ \cite{urushibara,horibaLSMO}) 
within the rigid-band model (i.e., assuming the 
constant exchange splitting) 
and that determined experimentally from core-level
photoemission spectra \cite{horibaLSMO}.
The experimental chemical potential shift is slower
than the calculated one by a factor of $\sim 0.4$, which means
\begin{equation}
\frac{\partial \mu}{\partial n}\sim \frac{0.4}{D^b(E_F)}.
\label{04}
\end{equation}
For a metallic system, the rigid-band picture predicts that
\begin{equation}
\frac{\partial \mu}{\partial n}=\frac{1+F}{D^{\ast}(E_F)}
=\left(\frac{m_b}{m^{\ast}}\right)\frac{1+F}{D^b(E_F)},
\label{F}
\end{equation}
where $D^{\ast}(E_F)$ is the DOS of renormalized quasiparticles (QPs)
and $F$ is a parameter which represents the effective QP-QP repulsion.
From Eqs.~(\ref{25})-(\ref{F}), one obtains $F\sim 0-0.4$.
This value is much smaller than the values of $F_0^s$ in AF compounds. 
For example, it was reported that
$F_0^s\sim 7$ for La$_{2-x}$Sr$_x$CuO$_4$, and $\sim 6$ for
La$_{1-x}$Sr$_x$VO$_3$ \cite{pote}. 
In the FM case, one
obtains small values of $F$ compared to 
the AF case, 
probably because of the strong exchange interaction between 
carriers in the FM case.

\begin{figure}
\begin{center}
\includegraphics[width=15cm]{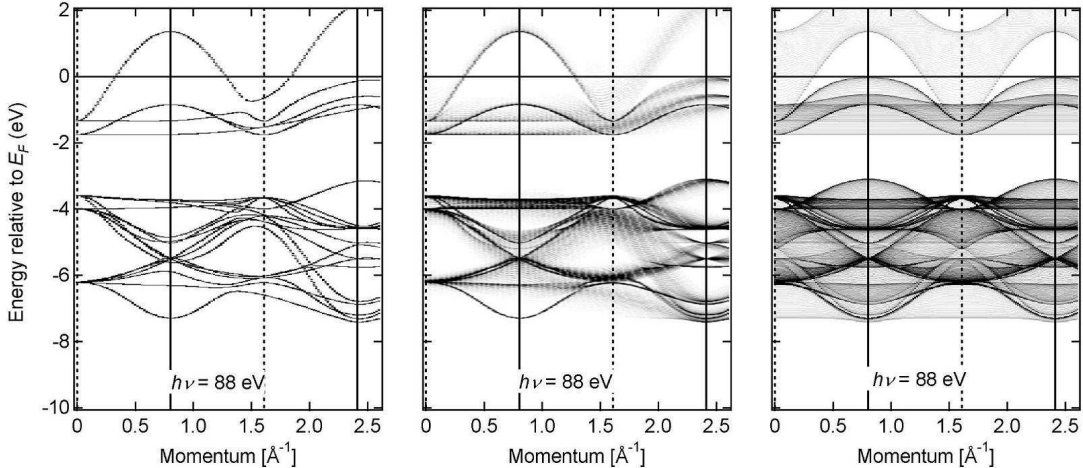}
\caption{\label{tight4} Simulated ARPES spectra of LSMO ($x=0.4$)
taken at 88 eV on TB calculation.
(a) $\Delta k_z = 0$, (b) $\Delta k_z =1/\lambda$,
(c) $\Delta k_z = \infty$.}
\end{center}
\end{figure}

Now we go back to the effect of the deviation of the trace
with $h\nu =88$ eV from the $\Gamma$-X direction.
Figure \ref{tight4}
shows a TB simulation of
the ARPES spectra of LSMO
taken at 88 eV using the above-mentioned
parameter values.
Panels (a), (b) and (c) show the results
for $\Delta k_z=0$ (without $k_z$ broadening),
$\Delta k_z=1/\lambda$,
and $\Delta k_z=\infty$
(uniform integration in $k_z$),
respectively. 
In panel (a), the energy position of
the bottom of the $e_g$ bands at
$k_{\parallel}=0 \ \mbox{\AA}^{-1}$ is
different from that at
$k_{\parallel}=1.6 \ \mbox{\AA}^{-1}$,
while in panel (b) the bottoms are
at almost the same energy positions. 
In experiment [Fig.~\ref{tight3} (a)], 
the bottoms are almost at the same 
energy positions, which means 
that the experimental results are in
good agreement with
(b), that is $\Delta k_z=1/\lambda$. In panel (c), there are two bands approaching
$E_F$, in disagreement with experiment.

\begin{figure}
\begin{center}
\includegraphics[width=6cm]{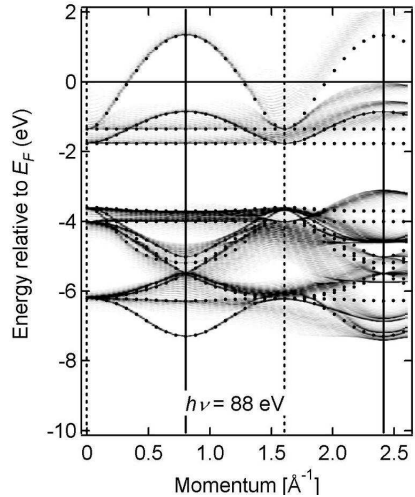}
\caption{\label{tight5} Comparison of the
dispersion along the $\Gamma$-X direction (dots)
and simulated spectra (gray scale) for LSMO ($x=0.4$)
calculated assuming $\Delta k_z =1/\lambda$ at 88 eV.}
\end{center}
\end{figure}

Figure \ref{tight5} shows comparison of the
dispersion along the $\Gamma$-X direction (dots)
and simulated spectra (gray scale) 
obtained assuming $\Delta k_z =1/\lambda$ for $h\nu = 88$ 
eV. 
There are some discrepancies in the high-binding-energy region
$-(4-7)$ eV, but the overall agreement is rather good
in the region near $E_F$ [$-(0-2)$ eV].
This means that one can trace the band dispersions almost
along
the $\Gamma$-X direction by performing ARPES with a
fixed photon energy of $h\nu=88$ eV 
because of the large $k_z$-integrated 
DOS at a high-symmetry line
(in this case the $\Gamma$-X direction).
Even if the trace somewhat deviates
from a high-symmetry line, one can trace the dispersion along
the high-symmetry line due to the large DOS and the effect of $k_z$
broadening.
By changing the photon energy with changing
the emission angle in ARPES measurements, one
can map band dispersions precisely along the
$\Gamma$-X line.
However,
the photoionization
cross section varies with
a photon energy, making the interpretation difficult.
In practice, fixing a photon energy and changing
an emission angle is a more efficient
and useful way to obtain band dispersions
in such materials. We also revealed
that considering the effect of $k_z$ broadening
is important for the interpretation of the ARPES of
the three-dimensional materials.

\section{Conclusion}
We have analyzed the ARPES data
of SVO bulk crystals
and LSFO ($x=0.4$) and LSMO
($x=0.4$) thin films using TB 
band-structure calculations. 
As for SVO, 
the enhanced effective electron mass 
obtained from the energy band near $E_F$ was consistent 
with the bulk thermodynamic properties. 
The experimental band structures of LSFO and LSMO
were interpreted
using a TB band-structure
calculation. The overall agreement between experiment and
calculation is fairly good.
However, in the case of LSFO, the energy position of
the calculated $E_F$ was not in agreement with the experimental $E_F$,
reflecting the insulating behavior. In the case of LSMO,
experimental energy bands were
narrowed compared with calculation
due to strong electron correlation.
We also demonstrated that considering the effect of
$k_z$ broadening is important for the interpretation
of the ARPES results of such three-dimensional 
materials.

\section*{Acknowledgments}
This work was supported by a 
Grant-in-Aid for Scientific Research No. A16204024 
from the Japan Society for the Promotion of Science (JSPS) 
and ``Invention of Anomalous Quantum
Materials'' No. 16076208 from the Ministry of Education, 
Culture, Sports, Science and Technology, Japan. 
The authors are grateful to D. H. Lu for technical 
support at SSRL. SSRL is operated 
by the Department of Energy's Office of Basic Energy
Science, Division of Chemical Sciences and Material 
Sciences. 
Part of this work was done under the approval 
of the Photon Factory Program Advisory Committee 
under Project No. 2002S2-002 at 
the Institute of Material Structure Science, KEK. 
H.W. acknowledges financial support from JSPS. 

\appendix
\section{Basis functions and matrix elements for 
TB band-structure calculations}
\begin{longtable}{ccccccc}
\caption{TB basis functions
used in the band-structure calculation of LSFO}\\
\toprule
 & No. & Origin & Function & No. & Origin & Function\\
\colrule
 Fe $3d$ & 1 & $(0,0,0)$ & $xy$ & 15 & $a(1,0,0)$ & $xy$\\
 & 2 & $(0,0,0)$ & $yz$ & 16 & $a(1,0,0)$ & $yz$\\
 & 3 & $(0,0,0)$ & $zx$ & 17 & $a(1,0,0)$ & $zx$\\
 & 4 & $(0,0,0)$ & $3z^2-r^2$ & 18 & $a(1,0,0)$ & $3z^2-r^2$\\
 & 5 & $(0,0,0)$ & $x^2-y^2$ & 19 & $a(1,0,0)$ & $x^2-y^2$\\
 O $2p$ & 6 & $\frac{1}{2}a(1,0,0)$ & $x$ & 20 & $\frac{1}{2}a(3,0,0)$ & $x$\\
 & 7 & $\frac{1}{2}a(1,0,0)$ & $y$ & 21 & $\frac{1}{2}a(3,0,0)$ & $y$\\
 & 8 & $\frac{1}{2}a(1,0,0)$ & $z$ & 22 & $\frac{1}{2}a(3,0,0)$ & $z$\\
 & 9 & $\frac{1}{2}a(0,1,0)$ & $x$ & 23 & $\frac{1}{2}a(2,1,0)$ & $x$\\
 & 10 & $\frac{1}{2}a(0,1,0)$ & $y$ & 24 & $\frac{1}{2}a(2,1,0)$ & $y$\\
 & 11 & $\frac{1}{2}a(0,1,0)$ & $z$ & 25 & $\frac{1}{2}a(2,1,0)$ & $z$\\
 & 12 & $\frac{1}{2}a(0,0,1)$ & $x$ & 26 & $\frac{1}{2}a(2,0,1)$ & $x$\\
 & 13 & $\frac{1}{2}a(0,0,1)$ & $y$ & 27 & $\frac{1}{2}a(2,0,1)$ & $y$\\
 & 14 & $\frac{1}{2}a(0,0,1)$ & $z$ & 28 & $\frac{1}{2}a(2,0,1)$ & $z$\\
\botrule
\label{symbol1}
\end{longtable}

\begin{longtable}{c}
\caption{Nonzero matrix elements of the TB Hamiltonian 
used in the calculation of LSFO. 
$\xi \equiv k_xa$, $\eta \equiv k_ya$, and $\zeta \equiv k_za$.}\\
\toprule
A. $d$-$d$ interactions\\
$H_{1,1}=H_{2,2}=H_{3,3}=\epsilon_{d}-4Dq-\Delta E/2$, 
$H_{4,4}=H_{5,5}=\epsilon_{d}+6Dq-\Delta E/2$ \\
$H_{15,15}=H_{16,16}=H_{17,17}=\epsilon_d-4Dq+\Delta E/2$,
$H_{18,18}=H_{19,19}=\epsilon_d+6Dq+\Delta E/2$\\
 \\
B. $p$-$p$ interactions\\
$C_1=\frac{1}{2}(pp\sigma)+\frac{1}{2}(pp\pi)$,
$C_2=\frac{1}{2}(pp\sigma)-\frac{1}{2}(pp\pi)$,
$C_3=(pp\pi)$\\
$H_{6,6}=H_{10,10}=H_{14,14}
=H_{20,20}=H_{24,24}=H_{28,28}=\epsilon_{p}$\\
$H_{7,7}=H_{8,8}=H_{9,9}
=H_{11,11}=H_{12,12}=H_{13,13}=\epsilon_{p}$\\
$H_{21,21}=H_{22,22}=H_{23,23}
=H_{25,25}=H_{26,26}=H_{27,27}=\epsilon_{p}$\\
$H_{6,9}=H_{7,10}=H_{20,23}=
H_{21,24}=2C_1\cos\left(\frac{1}{2}\xi-\frac{1}{2}\eta\right)$\\
$H_{6,10}=H_{7,9}=H_{20,24}=H_{21,23}=
-2C_2\cos\left(\frac{1}{2}\xi-\frac{1}{2}\eta\right)$\\
$H_{6,12}=H_{8,14}=H_{20,26}=H_{22,28}=
2C_1\cos\left(\frac{1}{2}\xi-\frac{1}{2}\zeta\right)$\\
$H_{6,14}=H_{8,12}=H_{20,28}=H_{22,26}=
-2C_2\cos\left(\frac{1}{2}\xi-\frac{1}{2}\zeta\right)$\\
$H_{6,23}=H_{7,24}=H_{20,9}=H_{21,10}=
2C_1\cos\left(\frac{1}{2}\xi+\frac{1}{2}\eta\right)$\\
$H_{6,24}=H_{7,23}=H_{20,10}=H_{21,9}=
2C_2\cos\left(\frac{1}{2}\xi+\frac{1}{2}\eta\right)$\\
$H_{6,26}=H_{8,28}=H_{20,12}=H_{22,14}=
2C_1\cos\left(\frac{1}{2}\xi+\frac{1}{2}\zeta\right)$\\
$H_{6,28}=H_{8,26}=H_{20,14}=H_{22,12}=
2C_2\cos\left(\frac{1}{2}\xi+\frac{1}{2}\zeta\right)$\\
$H_{7,13}=H_{21,27}=
2C_3\cos\left(\frac{1}{2}\xi-\frac{1}{2}\zeta\right)$\\
$H_{7,27}=H_{21,13}=
2C_3\cos\left(\frac{1}{2}\xi+\frac{1}{2}\zeta\right)$\\
$H_{8,11}=H_{22,25}=
2C_3\cos\left(\frac{1}{2}\xi-\frac{1}{2}\eta\right)$\\
$H_{8,25}=H_{22,11}=
2C_3\cos\left(\frac{1}{2}\xi+\frac{1}{2}\eta\right)$\\
$H_{9,12}=H_{23,26}=
2C_3\cos\left(\frac{1}{2}\eta-\frac{1}{2}\zeta\right)$\\
$H_{9,26}=H_{23,12}=
2C_3\cos\left(\frac{1}{2}\eta+\frac{1}{2}\zeta\right)$\\
$H_{10,13}=H_{11,14}=H_{24,27}=H_{25,28}=
2C_1\cos\left(\frac{1}{2}\eta-\frac{1}{2}\zeta\right)$\\
$H_{10,14}=H_{11,13}=H_{24,28}=H_{25,27}=
-2C_2\cos\left(\frac{1}{2}\eta-\frac{1}{2}\zeta\right)$\\
$H_{10,27}=H_{11,28}=H_{24,13}=H_{25,14}=
2C_1\cos\left(\frac{1}{2}\eta+\frac{1}{2}\zeta\right)$\\
$H_{10,28}=H_{11,27}=H_{24,14}=H_{25,13}=
2C_2\cos\left(\frac{1}{2}\eta+\frac{1}{2}\zeta\right)$\\
 \\
C. $p$-$d$ interactions\\
$H_{1,7}=H_{3,8}=H_{15,21}=H_{17,22}=
-(pd\pi)\exp\left(i\frac{1}{2}\xi\right)$\\
$H_{1,9}=H_{2,11}=H_{15,23}=H_{16,25}=
-(pd\pi)\exp\left(i\frac{1}{2}\eta\right)$\\
$H_{1,21}=H_{3,22}=H_{15,7}=H_{17,8}=
(pd\pi)\exp\left(-i\frac{1}{2}\xi\right)$\\
$H_{1,23}=H_{2,25}=H_{15,9}=H_{16,11}=
(pd\pi)\exp\left(-i\frac{1}{2}\eta\right)$\\
$H_{2,13}=H_{3,12}=H_{16,27}=H_{17,26}=
-(pd\pi)\exp\left(i\frac{1}{2}\zeta\right)$\\
$H_{2,27}=H_{3,26}=H_{16,13}=H_{17,12}=
(pd\pi)\exp\left(-i\frac{1}{2}\zeta\right)$\\
$H_{4,6}=H_{18,20}=
\frac{1}{2}(pd\sigma)\exp\left(i\frac{1}{2}\xi\right)$\\
$H_{4,10}=H_{18,24}=
\frac{1}{2}(pd\sigma)\exp\left(i\frac{1}{2}\eta\right)$\\
$H_{4,14}=H_{18,28}=
-(pd\sigma)\exp\left(i\frac{1}{2}\zeta\right)$\\
$H_{4,20}=H_{18,6}=
-\frac{1}{2}(pd\sigma)\exp\left(-i\frac{1}{2}\xi\right)$\\
$H_{4,24}=H_{18,10}=
-\frac{1}{2}(pd\sigma)\exp\left(-i\frac{1}{2}\eta\right)$\\
$H_{4,28}=H_{18,14}=
(pd\sigma)\exp\left(-i\frac{1}{2}\zeta\right)$\\
$H_{5,6}=H_{19,20}=
-\frac{\sqrt{3}}{2}(pd\sigma)\exp\left(i\frac{1}{2}\xi\right)$\\
$H_{5,10}=H_{19,24}=
\frac{\sqrt{3}}{2}(pd\sigma)\exp\left(i\frac{1}{2}\eta\right)$\\
$H_{5,20}=H_{19,6}=
\frac{\sqrt{3}}{2}(pd\sigma)\exp\left(-i\frac{1}{2}\xi\right)$\\
$H_{5,24}=H_{19,10}=
-\frac{\sqrt{3}}{2}(pd\sigma)\exp\left(-i\frac{1}{2}\eta\right)$\\
\botrule
\label{symbol2}
  \end{longtable}

\begin{longtable}{cccccccc}
\caption{TB basis functions
used in the band-structure calculation of LSMO}\\
\toprule
 & No. & Origin & Function &  & No. & Origin & Function \\
\colrule
 Fe $3d$ & 1 & $(0,0,0)$ & $xy$ &  O $2p$ & 6 & $\frac{1}{2}a(1,0,0)$ & $x$\\
 & 2 & $(0,0,0)$ & $yz$ & & 7 & $\frac{1}{2}a(1,0,0)$ & $y$ \\
 & 3 & $(0,0,0)$ & $zx$ & & 8 & $\frac{1}{2}a(1,0,0)$ & $z$ \\
 & 4 & $(0,0,0)$ & $3z^2-r^2$ & & 9 & $\frac{1}{2}a(0,1,0)$ & $x$\\
 & 5 & $(0,0,0)$ & $x^2-y^2$ & & 10 & $\frac{1}{2}a(0,1,0)$ & $y$\\
 & & & & & 11 & $\frac{1}{2}a(0,1,0)$ & $z$ \\
 & & & & & 12 & $\frac{1}{2}a(0,0,1)$ & $x$ \\
 & & & & & 13 & $\frac{1}{2}a(0,0,1)$ & $y$ \\
 & & & & & 14 & $\frac{1}{2}a(0,0,1)$ & $z$ \\
\botrule
\label{symbol3}
\end{longtable}

\begin{longtable}{c}
\caption{Nonzero matrix elements of the TB Hamiltonian 
used in the calculation of LSMO. 
$\xi \equiv k_xa$, $\eta \equiv k_ya$, and $\zeta \equiv k_za$.}\\
\toprule
A. $d$-$d$ interactions\\
$H_{1,1}=H_{2,2}=H_{3,3}=\epsilon_d-4Dq-\Delta E/2$ (up spin), 
$\epsilon_d-4Dq+\Delta E/2$ (down spin)\\
$H_{4,4}=H_{5,5}=\epsilon_d+6Dq-\Delta E/2$ (up spin), 
$\epsilon_d+6Dq+\Delta E/2$ (down spin)\\
 \\
B. $p$-$p$ interactions\\
$C_1=\frac{1}{2}(pp\sigma)+\frac{1}{2}(pp\pi)$,
$C_2=\frac{1}{2}(pp\sigma)-\frac{1}{2}(pp\pi)$,
$C_3=(pp\pi)$\\
$H_{6,6}=H_{10,10}=H_{14,14}=\epsilon_{p}$\\
$H_{7,7}=H_{8,8}=H_{9,9}=
H_{11,11}=H_{12,12}=H_{13,13}=
\epsilon_{p}$ \\
$H_{6,9}=H_{7,10}=4C_1\cos(\frac{1}{2}\xi)\cos(\frac{1}{2}\eta)$\\
$H_{6,10}=H_{7,9}=-4C_2\sin(\frac{1}{2}\xi)\sin(\frac{1}{2}\eta)$\\
$H_{6,12}=H_{8,14}=4C_1\cos(\frac{1}{2}\xi)\cos(\frac{1}{2}\zeta)$\\
$H_{6,14}=H_{8,12}=-4C_2\sin(\frac{1}{2}\xi)\sin(\frac{1}{2}\zeta)$\\
$H_{7,13} = 4C_3\cos(\frac{1}{2}\xi)\cos(\frac{1}{2}\zeta)$,
$H_{8,11} = 4C_3\cos(\frac{1}{2}\xi)\cos(\frac{1}{2}\eta)$\\
$H_{9,12} = 4C_3\cos(\frac{1}{2}\eta)\cos(\frac{1}{2}\zeta)$\\
$H_{10,13}=H_{11,14}=4C_1\cos(\frac{1}{2}\eta)\cos(\frac{1}{2}\zeta)$\\
$H_{10,14}=H_{11,13} -4C_2\sin(\frac{1}{2}\eta)\sin(\frac{1}{2}\zeta)$\\
\\
C. $p$-$d$ interactions\\
$H_{1,7}=H_{3,8}=-2i(pd\pi)\sin(\frac{1}{2}\xi)$\\
$H_{1,9}=H_{2,11}=-2i(pd\pi)\sin(\frac{1}{2}\eta)$\\
$H_{2,13}=H_{3,12}=-2i(pd\pi)\sin(\frac{1}{2}\zeta)$\\
$H_{4,6}=i(pd\sigma)\sin(\frac{1}{2}\xi)$,
$H_{4,10}=i(pd\sigma)\sin(\frac{1}{2}\eta)$\\
$H_{4,14}=-2i(pd\sigma)\sin(\frac{1}{2}\zeta)$,
$H_{5,6}=-\sqrt{3}i(pd\sigma)\sin(\frac{1}{2}\xi)$\\
$H_{5,10}=\sqrt{3}i(pd\sigma)\sin(\frac{1}{2}\eta)$\\
\botrule
\label{symbol4}
  \end{longtable}
\bibliography{LSFOtex,SVO}

\label{lastpage}

\end{document}